\documentstyle[12pt,epsfig]{article}
\newlength{\dinwidth}
\newlength{\dinmargin}
\newcommand{\ba}{\begin{array}}
\newcommand{\ea}{\end{array}}
\newcommand{\be}{\begin{equation}}
\newcommand{\ee}{\end{equation}}
\newcommand{\bea}{\begin{eqnarray}}
\newcommand{\eea}{\end{eqnarray}}
\newcommand{\gsim}{\mathrel{\mathop{\kern 0pt \rlap
  {\raise.2ex\hbox{$>$}}} \lower.9ex\hbox{\kern-.190em $\sim$}}}

\def\pz{\partial_z}
\def\pbz{\partial_{\overline z}}
\def\half{{1\over 2}}

\def\bz{{\overline z}}
\def\bh{{\overline h}}
\def\bL{{\overline L}}
\def\bU{{\overline U}}
\def\bV{{\overline V}}
\def\bj{{\overline j}}
\def\nn{\nonumber}
\def\cO{{\cal O}}

\def\bw{{\overline{w}}}
\def\btheta{{\overline{\theta}}}
\def\ba{{\overline a}}
\def\bb{{\overline b}}
\def\sa{{a^\star}}
\def\sb{{b^\star}}
\def\bsa{{\overline{a^\star}}}
\def\bsb{{\overline{b^\star}}}

\def\vx{{\vec{x}}}
\def\vy{\vec{y}}
\def\cO{{\cal{O}}}

\def\log{{\rm{log}}}

\def\br{{\overline r}}

\def\btheta{{\overline \theta}}
\def\bm{{\overline m}}
\def\bn{{\overline n}}

\def\bt{{\overline t}}
\def\cO{{\cal O}}

\def\bL{\overline L}
\def\bJ{\overline J}
\def\brho{\overline \rho}
\def\bA{\overline A}

\def\cn{{\cal N}}

\def\ben{\begin{equation}}
\def\een{\end{equation}}
\def\bea{\begin{eqnarray}}
\def\eea{\end{eqnarray}}

\begin{document}
\thispagestyle{empty}
\addtocounter{page}{-1}
\vskip-0.35cm
\begin{flushright}
UK/02-08 \\
TIFR-TH/02-19\\
IFT UAM-CSIC/02-19 \\
{\tt hep-th/0206062}
\end{flushright}
\vspace*{0.2cm}
\centerline{\Large \bf Realizations of Conformal and Heisenberg Algebras}
\vskip0.3cm
\centerline{\Large \bf in PP-wave-CFT correpondence}

\vspace*{1.0cm} 
\centerline{\bf Sumit R. Das${}^{a,b}$ and Cesar Gomez$^{c}$ }
\vspace*{0.7cm}
\centerline{\it Department of Physics and Astronomy,}
\vspace*{0.2cm}
\centerline{\it University of Kentucky, Lexington, KY 40506 \rm USA ${}^a$}
\vspace*{0.35cm}
\centerline{\it Tata Institute of Fundamental Research}
\vspace*{0.2cm}
\centerline{\it Homi Bhabha Road, Mumbai 400 005 \rm INDIA ${}^b$}
\vspace*{0.35cm}
\centerline{\it Instituto de Fisica Teorica, C-XVI Universidad Autonoma,}
\vspace*{0.2cm}
\centerline{\it E-28049 Madrid \rm SPAIN ${}^c$}
\vspace*{1cm}
\centerline{\tt das@theory.tifr.res.in 
\hskip0.7cm Cesar.Gomez@uam.es}

\vspace*{0.8cm}
\centerline{\bf Abstract}
\vspace*{0.3cm}

We elaborate on the symmetry breaking pattern involved in the Penrose
limit of $AdS_{d+1} \times S^{d+1}$ spacetimes and the corresponding limit
of the CFT dual. For $d=2$ we examine in detail how the symmetries
contract to products of rotation and Heisenberg algebras, both from
the bulk and CFT points of view. Using a free field realization of
these algebras acting on products of elementary fields of the
CFT with $SO(2)$ R
charge $+1$ , we show that this process of contraction restricts all
the fields to a few low angular momentum modes and ensures that the
field with R charge $-1$ does not appear. This provides an
understanding of several important aspects of the proposal of
Berenstein, Maldacena and Nastase. We also indicate how the contraction
can be performed on correlation functions.

\baselineskip=18pt

\newpage

\section{Introduction}

PP waves provide exact backgrounds for string theory in which the
Green-Schwarz worldsheet action becomes quadratic in the lightcone
gauge \cite{metsaev}. Such plane waves may be regarded as Penrose
limits \cite{penrose} of $AdS \times S$ spacetimes \cite{blau}.  Some
time ago, Berenstein, Maldacena and Nastase (BMN) \cite{bmn} used this
to propose that the dual description of IIB string theory on the ten
dimensional pp-wave is the large R-charge sector of the ${\cal
N}=4,~SU(N)$ gauge theory. Questions of holography in this background
appear to be confusing at this moment and various proposals have
been made in \cite{dgr,kp,lor,bnas,siopsis}.
The proposal has been extended to
other related backgrounds \cite{other}. Open strings and D-branes
have been studied in this context in \cite{dbranes}. Further
insight into the correspondence has been obtained from a
semiclassical treatment \cite{semiclassical}. Other aspects of string
theory in pp-wave background has been studied in \cite{strings}.
Properties of the Yang-Mills theory relevant to this limit have been
studied in \cite{larger}. Some questions about black hole formation
in these backgrpounds have been addressed in \cite{bh}.

In \cite{dgr}, the Penrose limit leading to a pp-wave
was considered from the point of view of symmetry breaking.
The bosonic isometries of $AdS_{d+1} \times S^{d+1}$
backgrounds are $SO(d,2) \times SO(d+2)$. In the pp wave limit the
number of Killing vectors remain the same and include $SO(d) \times
SO(d) \times H(d) \times H(d)$ together with translations along the
two light cone directions. In terms of the standard metric of a
pp-wave \ben ds^2 = 2dx^+ dx^- - \mu^2(\vx^2 + \vy^2)(dx^+)^2 + (d
\vx)^2 + (d\vy)^2
\label{eq:one}
\een the first $H(d)$ denotes a Heisenberg algebra which acts on the
plane $\vx$ transverse to the wave, while the second $H(d)$ acts on
the plane $\vy$. The two $SO(d)$ factors are rotations in the two
transverse planes $\vx$ and $\vy$. (The metric itself has more
symmetries, but the RR gauge fields which are necessary for the
solution do not).  However, the generators of $H(d)$ are {\em broken}
symmetries since they do not commute with the light cone time
$x^+$. Rather, $\partial_-$ is the common central charge for both the
Heisenberg algebras while $\partial_+$ acts as an outer automorphism.
The single particle states in this background may be then considered
as Nambu-Goldstone bosons. Other discussions of symmetries appear
in \cite{symmetries}.

In the bulk, supergravity states are created from the light cone
vacuum with a given value of the central charge by the action of the
creation operators of $H(d) \times H(d)$. This is similar to what
happens in $AdS \times S$ where the states are created by the raising
operators of the conformal isometries and those of $SO(d+2)$. For $AdS
\times S$, the $AdS$ symmetry generators simplify near the boundary
and become differential operators which act entirely on the $S^{d-1}
\times ({\tt time})$. These then become the conformal symmetries of
the holographic theory defined on the boundary, while the $SO(d+2)$
acts as internal symmetries. These symmetries are in turn used to
create states in the CFT which are dual to the supergravity modes.

In the Penrose limit, we
have to focus on the {\em center} of $AdS$ rather than the boundary.
In this limit the $AdS \times S$ isometries become the pp-wave
isometries which include the Heisenberg algebras. The latter act on
the transverse planes while translation in $x^+$ generate identical
copies of the same algebra.  In \cite{dgr} this was used to suggest
that there should be a holographic description in which these
symmetries are realized on a $d$ dimensional transverse plane and
$x^+$ acts as a holographic coordinate representing a scale. The other
half of these symmetries would be still realized as internal
symmetries. Other authors \cite{kp} proposed that the holographic
theory should involve the entire transverse plane.  In \cite{lor} the
radial coordinate of the transverse plane was proposed as a
holographic direction. In \cite{bnas} it was shown that the boundary
of the pp wave spaetime is a one dimensional null line and it was
suggested that this would be the place where a holographic theory
should live.

These observations seem to be contradict the original proposal that
the dual theory is the same old Yang-Mills theory restricted to the
large R-charge sector, since this gauge theory clearly lives on the
boundary of $AdS_{d+1} \times S^{d+1}$ which is $S^{d-1} \times {\tt
time}$. In fact, this boundary is not a part of the pp-wave
spacetime. While it is certainly true that this sector of the
Yang-Mills reproduces quantities in the string theory, this is not a
holographic description in the usual sense.

In this note we shed light on this confusing issue by analyzing how
the original symmetries of the theory "contract" in the Penrose limit,
both from the bulk point of view as well as from the gauge theory
point of view. 

For the simple case of $d=2$ we first show how Heisenberg algebras
arise from the conformal and rotation algebras at the abstract level.
We then study in detail a free field realization involving two complex
scalar fields. This serves as a toy model for more realistic cases as
in e.g. $d=4$.  Starting with a highest weight state created by
products of the complex scalar which has $SO(2)$ R charge $+1$ and
creating states by lowering operators of the conformal and rotation
groups, we show explicitly how for large conformal weights and R charges
the states organize as representations of Heisenberg
algebras. We find that at the same time the fields become restricted
to a few low angular momentum modes and the scalar with R charge $-1$
never appears. We indicate how the result would generalize to
$AdS_{d+1} \times S^{d+1}$ with $d > 2$, though we do not work out the
details. In these cases the reduction of the number of modes is in
fact expected to be simpler : the field with $SO(2)$ R charge $+1$ is
restricted to the zero mode and $d$ other lowest angular momentum
modes, the fields neutral under the $SO(2)$ are in their zero modes
while the field with $SO(2)$ R charge $-1$ does not participate. These
are crucial ingredients of the proposal of BMN where it was argued
that operators which contain modes other than those listed above would
generically have large anomalous dimensions in large $N$. Here we have
shown that the same restrictions are required by the contraction of
the conformal and rotation algebras to the Heisenberg algebras.
Finally we discuss a way to redefine correlation functions in the CFT
appropriate to the large weight, large R-charge limit.

\section{Isometries in $AdS$ and the $AdS/CFT$ correspondence}

Let us recall some aspects of the standard $AdS/CFT$ correspondence
\cite{adscft}
relevant to our discussion. We will explicitly deal with $AdS_3 \times
S^3$ for simplicity. The results generalize to other dimensions in a
straightforward way which we will indicate. In global coordinates the
metric is 
\ben ds^2 = R^2 [ -(1+\br^2)d\bt^2 + {d\br^2 \over 1 +
\br^2} + \br^2 d\chi^2 + (1- \brho^2)d\btheta^2 + {d\brho^2 \over 1-
\brho^2} + \brho^2 d\phi^2]
\label{eq:two}
\een 
The isometries are $SL(2,R) \times SL(2,R) \times SU(2) \times
SU(2)$ where the two $SL(2,R)$ factors come from the $AdS_3$ part and
the two $SU(2)$ factors are the standard symmetries of the $S^3$. We
will denote these generators as 
\ben L_0, L_-, L_+;~~~\bL_0, \bL_-,
\bL_+;~~~~J_0,J_-,J_+;~~~~\bJ_0, \bJ_-, \bJ_+
\label{eq:three}
\een
In terms of null coordinates 
\ben
w = t + \chi~~~~~~~~~~\bw = t-\chi
\label{eq:four}
\een
the $L_i$ are \cite{bala}
\bea
L_0 & = & i\partial_w \nn \\
L_- & = & i e^{-iw}[{2\br^2 + 1 \over 2\br {\sqrt{1 + \br^2}}} \partial_w - 
{1 \over 2\br {\sqrt{1+\br^2}}} \partial_\bw +
{i\over 2}{\sqrt{1+\br^2}}\partial_\br] \nn \\
L_+ & = & ie^{iw}[{2\br^2 + 1 \over 2\br {\sqrt{1 + \br^2}}} \partial_w - 
{1 \over 2\br {\sqrt{1+\br^2}}} \partial_\bw -
{i\over 2}{\sqrt{1+\br^2}}\partial_\br] 
\label{eq:five}
\eea
These satisfy the commutation relations
\ben
[ L_0,  L_\pm ] = \mp L_\pm~~~~~~~~[L_+, L_-] = 2 L_0
\label{eq:six}
\een
The expressions for $\bL_i$ are obtained by interchanging $w$ and
$\bw$. The expressions for $J_i$ and $\bJ_i$ can be similalrly written
down by analytically continuing these expressions. We will not need
these, since (\ref{eq:five}) would be sufficient to make the point.

The states of some field in the bulk form some representation of the
algebra.  Consider for example a massless scalar field in six
dimensions. The lowest energy state is a highest weight state which
satisfies 
\ben
L_+ ~|h> = \bL_+ ~|h> = 0~~~~~~L_0 ~|h> = \bL_0 ~|h> = h
~|h>
\label{eq:seven}
\een 
The weight $h$ is determined in terms of the $SO(4) = SU(2)
\times SU(2) $ representation content. The creation of states in each
represntation is standard and will not be repeated here. If the
$SO(4)$ angular momentum is $L$, so that the quadratic Casimir is
$L(L+2)$ one has 
\ben 
2h = L + 2
\label{eq:eight}
\een
The descendants may be obtained by the action of the $L_-$'s
\ben
~|n,\bn,h> = (L_-)^n (L_-)^\bn ~|h>
\label{eq:nine}
\een
This has
\ben
L_0 ~|n,\bn,h> = (h+n) ~|n,\bn,h> ~~~~~\bL_0 ~|n,\bn,h> = (h + \bn)~|n,\bn,h>
\label{eq:ten}
\een
As is clear from the expressions for the $L_0$ and $\bL_0$, the total
energy $\omega$ is the value of $L_0 + \bL_0$ while the angular momentum 
$l$ along
the circle whose coordinate is $\chi$ is the value of $\bL_0 -
L_0$. Thus we have 
\ben \omega = 2h + n + \bn~~~~~~~~~~~~l = \bn- n
\label{eq:eleven}
\een

It will be useful to relate this discussion to the normalizable
solutions of the wave equation. The solution for the quantum numbers
given above is given by
\ben
\Psi (\bt,\br,\chi;\Omega) = e^{-i\omega \bt} Y_L (\Omega)~e^{il\chi}
 [{1\over {\sqrt{1+\br^2}}}]^{2h}~[{\br \over {\sqrt{1+\br^2}}}]^l ~
F(2h+n+l,-n;2h;{1 \over {\sqrt{1+\br^2}}})
\label{eq:twelve}
\een
Using the properties of hypergeometric functions this may be rewritten as
\ben
\Psi (\bt,\br,\chi;\Omega) = e^{-i\omega \bt} Y_L (\Omega)~e^{il\chi}
 [{1\over {\sqrt{1+\br^2}}}]^{2h}~[{\br \over {\sqrt{1+\br^2}}}]^l ~
F(2h+n+l,-n;l+1;{\br^2 \over {\sqrt{1+\br^2}}})
\label{eq:twelvea}
\een
The key property which leads to a holographic description on the
boundary at $\br = \br_0 \rightarrow \infty$ is that these wave
functions have a universal behavior for large $\br$.  It is clear from
(\ref{eq:twelve}) that $\Psi \sim (\br_0)^{-2h}$ regardless of the
values of $n$ and $l$. This means that a local bulk operator becomes a
local boundary operator up to an overall factor of the cutoff
$\br_0$. In the holographic description it is this boundary operator
which creates the state and $\br_0$ becomes a scale in the theory.

This behavior of the wavefunction also reduces the generators of $L_i$ and
$\bL_i$ to standard forms. On the boundary one gets
\bea
L_0 & = & i\partial_w \nn \\
L_- & = & ie^{-iw}[\partial_w - ih] \nn \\
L_+ & = & ie^{iw}[\partial_w + ih]
\label{eq:thirteen}
\eea
which are the standard form of $SL(2,R)$ generators acting on a
primary field of weight $h$. There are similar expressions for
$\bL_i$. Thus the states can be created in the CFT side in a fashion
identical to that in the bulk. The difference is that now the
wavefunctions are functions of $w,\bw$, i.e. in a $1+1$ dimensional
theory on the boundary. The generators of the $S^3$ isometries remain
the same as we approach the boundary since they do not involve
$\br$. These symmetries are realized as internal symmetries in the
CFT. This makes it clear why the natural location of the holographic
theory is on the boundary.

The above discussion may be easily generalized to higher
dimensions. Representing $AdS_{d+1}$ by an equation
\ben
X_1^2 + X_2^2 - X_3^2 - \cdots X_{d+2}^2 = R^2
\label{eq:fourteen}
\een
in a flat $d+2$ dimensional space with signature $(-1,-1,1,1,\cdots)$
the conformal isometries are rotations in this embedding space,
denoted by $J_{AB}$.  Writing the $AdS_{d+1}$ metric as
\ben
ds^2 = R^2[-(1+\br^2)d\bt^2 + {d\br^2 \over 1 + \br^2} + \br^2 d\Omega^2]
\label{eq:fifteen}
\een
where $d\Omega^2$ is the standard metric on a $d-1$ dimensional
sphere, it is clear that the generators $J_{ij}, i,j = 3 \cdots d+2$
are the $SO(d)$ rotations on this sphere. The energy is given by the
generator $J_{12}$. These are then the generalizations of the
generators $L_0$ and $\bL_0$ of the three dimensional case. The
remaining generators $J_{1i}$ and $J_{2i}$ are the conformal
symmetries which are the generalizations of $L_\pm$ and
$\bL_\pm$. Once again these become the standard generators of
conformal symmetries on the boundary.

\section{The Penrose limit and PP waves}

To get to the six dimensional pp-wave from $AdS_3 \times S^3$, we first define
\bea
t & = & \bt \cosh \alpha - \btheta \sinh \alpha \nn \\
\theta & = & \bt \sinh \alpha + \btheta \cosh \alpha
\label{eq:bone}
\eea
Then we rescale
\ben
r = R \br ~~~~~~\rho = R \rho~~~~~~~ x^\pm = {R \over {\sqrt{2}}}
(\theta \pm t)
\een
and take the limit
\ben
R,\alpha \rightarrow \infty~~~~~~~r,\rho,x^\pm = {\rm fixed}~~~~~~~~
\mu = {e^\alpha \over {\sqrt {2}} ~R} = {\rm fixed}
\een
In this limit the metric reduces to (\ref{eq:one}) with the definitions
\ben
\vx = (x_1,x_2)~~~~~~~x_1 = r \cos \chi~~~~~ x_2 = r \sin \chi
\een
\ben
\vy = (y_1,y_2)~~~~~~~y_1 = \rho \cos \phi~~~~~~~~y_2 = \rho \sin \phi
\een
The parameter $\mu$ may be set to unity (if nonzero) by rescaling
$x^\pm$. In the following we will set $\mu = 1$. These formulae can be
trivially generalized to higher dimensions.

We want to see what happens to the symmetry generators of $AdS_3
\times S^3$ in this limit. This means - among other things - that we
have to take the $SL(2,R)$ generators given in (\ref{eq:five}) and
focus on the region $\br \rightarrow 0$ keeping $r$ defined above
fixed. This is the opposite of the limit taken to reduce the
isometries of the bulk to conformal symmetries on the
boundary. Performing the Penrose limit as described above we get the
following limiting form of the generators
\bea
L_0 + \bL_0 & = & i\partial_+ - 2i R^2 \partial_- \nn \\
L_0 - \bL_0 & = & i \partial_\chi \nn \\
L_- + \bL_- & = & R e^{-ix^+}[{\partial \over \partial x_1} 
-ix_1 \partial_-] \equiv R a_1^\dagger \nn \\
L_- - \bL_- & = & i R e^{-ix^+}[{\partial \over \partial x_2} 
-ix_2 \partial_-] \equiv R a_2^\dagger \nn \\
L_+ + \bL_+ & = & R e^{ix^+}[{\partial \over \partial x_1} 
+ ix_1 \partial_-] \equiv R a_1 \nn \\
L_- + \bL_- & = & R e^{ix^+}[{\partial \over \partial x_2} 
+ ix_2 \partial_-] \equiv R a_2 
\label{eq:bthree}
\eea 
In this limit the algebra becomes 
\ben
[ a_i, a_j^\dagger ] =  -2i \delta_{ij} \partial_-
\een
while
\ben
[ L_0 - \bL_0 , a_1 ]  =  a_2~~~~~~~~~[ L_0 - \bL_0, a_2 ] = - a_1
\label{eq:bfour}
\een
The generator $\partial_-$ commutes with every other generator and
acts as a central charge. To leading order this is essentially $L_0 +
\bL_0$.  The $a_j, a_j^\dagger$ then form a Heisenberg algebra.  The
limiting form of the generators in (\ref{eq:bthree}) are well known
symmetries of the pp-wave background \cite{symmetries}.

From the isometries of $S^3$ we get a similar structure, viz two more
sets of oscillators $b_1,b_1^\dagger,b_2,b_2^\dagger$ which may be
obtained by replacing $x_j$ in ({\ref{eq:bthree}) by the $y_j$. In
addition we have
\bea
J_0 + \bJ_0 & = & -i\partial_+ - i R^2 \partial_- \nn \\
J_0 - \bJ_0 & = & i \partial_\phi
\label{eq:bfive}
\eea
Significantly, the central charge which appears in the oscillator
algebra $b_j$ is the same as that in the $a_j$ algebra.

The element $\partial_+$ acts as an outer automorphism, generating
identical copies of the algebra at different values of $x^+$. From the
expressions above we have
\bea
L_0 +\bL_0 - J_0 - \bJ_0 & = & i\partial_+ \nn \\
L_0 + \bL_0 + J_0 + \bJ_0 & = & -iR^2 \partial_-
\label{eq:bsix}
\eea

To obtain the supergravity 
states of the bulk theory we start with the "light cone
vacuum" specified by the value of the central charge $-i\partial_- =
p_-$ which is annihilated by all the destruction operators $a_j,
b_j$. This is the lowest energy state. The higher states are created
as usual by action of the creation operators just as in a
multidimensional harmonic oscillator. Thus one has states of the form
\ben 
~|n_i, m_a, p_-> = \prod_i (a_i^\dagger)^{n_i} \prod_a
(b_a^\dagger)^{m_a}~|0,p_->
\label{eq:bseven}
\een
The light cone energy is given by
\ben
p^- \equiv p_+ = \sum_j n_j + \sum_a m_a + 2
\label{eq:beight}
\een
The wave functions are given by standard Hermite polynomials in $x_j,y_a$.

It is interesting to see how the $AdS$ wavefunctions become these wave
functions. To do that it is easier to work in the coordinates
$r,\rho,\chi,\phi$. Consider a massless scalar in six dimensions. The
wavefunctions are then 
\ben 
\Psi = e^{-ip_+x^+ + ip_-x^-}e^{il\chi + ij
\phi}e^{-\half p_-(r^2+\rho^2)}~L_n^l(p_-r^2)~L_m^j (p_-\rho^2)
\label{eq:bnine}
\een
where $L$ denotes a Laguerre function. The dispersion relation is then 
\ben
p_+ = 2n+l + 2m + j + 2
\label{eq:bten}
\een 
Equation (\ref{eq:bnine}) should be compared with (\ref{eq:twelvea})
in the Penrose limit \footnote{ A similar comparison was performed in
\cite{lor}, however the wavefunctions used in this paper did not
contain the Laguerre polynomial piece.}.  First consider the radial part
of (\ref{eq:twelvea}). For large angular momentum on the $S^3$ we have
$2h \sim p_-R^2$. Near $\br = 0$ with $r = R\br$ fixed the first two
factors become \ben [{1\over {\sqrt{1+\br^2}}}]^{2h}~[{\br \over
{\sqrt{1+\br^2}}}]^l \sim e^{-\half p_- r^2}~r^l
\label{eq:beleven}
\een
while the hypergeometric function simplifies to \footnote{Note that since
the third argument of the hypergeometric function is a negative integer,
it is a polynomial rather than an infinite series.} 
\ben
F(p_-R^2,-n;l+1;\br^2) \sim L_n^l(p_-r^2)
\label{eq:btwelve}
\een
The expressions (\ref{eq:btwelve}) and (\ref{eq:beleven}) then gives the $r$
dependent part of (\ref{eq:bnine}).
The spherical harmonic $Y_L$ may be written in the form
\ben
Y_L (\Omega) = e^{iJ\btheta}e^{ij \phi}
[{1\over {\sqrt{1-\brho^2}}}]^{-2h}~[{\brho \over {\sqrt{1-\brho^2}}}]^j ~
 F(2h+m+l,-m;j+1;{\brho^2 \over {\sqrt{1-\brho^2}}})
\label{eq:bthirteen}
\een 
This leads to the $\rho$ dependent part of
(\ref{eq:bnine}). Finally noting that $J = R^2 p_-$ the $\bt$ and
$\btheta$ factors of (\ref{eq:bthirteen}) and (\ref{eq:twelvea})
combine to give the $x^\pm$ dependent phases of (\ref{eq:bnine}).

Once again it is straightforward to generalize the discussion to
higher dimensions. As explained at the end of the previous section the
generators $J_{ij}$ of the $AdS$ part of the isometries remain as they
are and become the $SO(d)$ generators acting on the $\vx$ part of the
transverse plane. To leading order, the generator $J_{12} =
\partial_\bt$ becomes a central term, and the generators $J_{1i}$ and
$J_{2i}$ combine to form the Heisenberg algebra on the $\vx$ plane. On
the $S$ side, the story is similar. The $SO(d)$ rotations simply carry
over. The other off-diagonal generators reduce to Heisenberg algebra
generators acting on the transverse plane $\vy$.  The generator
$\partial_\btheta$ becomes a central term and is equal to
$\partial_\bt$ to leading order. The difference between $\partial_\bt$
and $\partial_\btheta$ is however finite and provides the outer
automorphism $\partial_+$ while the sum provides the common central
term of the two Heisenberg algebras. The reduction of the
wavefunctions to the correct pp-wave wavefunctions also follow.

The generators (\ref{eq:bthree}) involve derivatives with respect to
the radial coordinates $r$ and $\rho$ in an essential way. There is
no simple way in which these are related to the way the conformal
algebra acts on the dual theory via the generators in (\ref{eq:thirteen})
simply because the Penrose limit involved going to the $\br = 0$ region.
From this point of view it appears mysterious how one would realize the
algebra of symmetries of the pp-wave in the dual gauge theory which 
lives in a region which is {\em not contained in the pp-wave geometry}.

\section{From Conformal and Rotation algebras to Heisenberg algebras}

Before delving into the question how the pp-wave symmetries are realized
in the dual theory let us examine how the contraction of conformal and
rotation algebras happens at the abstract level. Specifically we will
show that the conformal algebra reduces to a product of rotation algebra
and a Heisenberg alegbra when we act on states with large conformal
weight. Similarly, when a rotation algebra acts on states of large
angular momentum, it reduces to a product of a lower dimensional 
rotation algebra and a Heisenberg algebra. Finally we will combine
the original conformal and rotation algebras. We will discuss the
case of $SO(2,2) \times SO(4)$ in detail. The result generalizes easily
to other $SO(d,2) \times SO(d+2)$.

\newpage

\subsection{Conformal algebra $\rightarrow$ Heisenberg algebra}

Each of the $SL(2,R)$ factors of $SO(2,2)$ satisfies
the algebra (\ref{eq:six}). We will consider a highest weight state
$~|h,\bh;0,0>$
\bea
L_+~|h,\bh;0,0> & = & \bL_+~|h,\bh;0,0> = 0 \nn \\
L_0 ~|h,\bh,0,0> & = & h ~|h,\bh;0,0> \nn \\
\bL_0 ~|h,\bh,0,0> & = & \bh ~|h,\bh,0,0>
\label{eq:cone}
\eea
and the descendants of level $(n,\bn)$ 
\ben
~|h,\bh,n,\bn> = (L_-)^n (\bL_-)^\bn ~|h,\bh,0,0>
\label{eq:ctwo}
\een
Consider first the action of one of the $SL(2,R)$'s. The algebra implies
\bea
L_0 ~|h,\bh,n,\bn> & = & (h+n) ~|h,\bh,n,\bn>~~~~~L_- ~|h,\bh,n,\bn> = 
~|h,\bh,n+1,\bn> \nn \\
L_+ ~|h,\bh,n,\bn> & = & [2nh + n(n-1)]~|h,\bh,n-1,\bn>
\label{eq:cthree}
\eea
It is now clear that when $n << h$ the generator $L_0$ becomes a
c-number, and 
\ben 
L_- ~|h,\bh,n,\bn> = ~|h,\bh,n+1,\bn>~~~~~~L_+ ~|h,\bh,n,\bn>
= 2nh~|h,\bh,n-1,\bn>
\label{eq:cfour}
\een
These are precisely the relations obtained from the action of
annihilation and creation operators of a Heisenberg algebra.
Thus $L_+$ and $L_-$ may be regarded as annihilation and creation
operators $A$ and $A^\dagger$ respectively with
\ben
[A, A^\dagger] = 2h
\label{eq:cfive}
\een
An identical contraction of course happens for the other $SL(2,R)$
piece.  Acting by generators $\bL_-$ we get states which are specified
by another level number $\bn$ and we have another Heisenberg algebra
$\bA, \bA^\dagger$ which commute to $2\bh$. The sum and difference
$L_+ \pm \bL_+$ behave as two independent annihilation operators
while $L_- \pm \bL_-$ behave as creation operators and these commute to
$2(h + \bh)$.
The {\em difference} $L_0 - \bL_0$
usual the angular momentum of the state. 
Thus we
have an algebra $SO(2) \times H(2) $ together with a central charge.

The above discussion may be easily generalized to $SO(d,2)$. The role
of $L_\pm, \bL_\pm$'s is now replaced by the $d$ sets of raising and
lowering operators. The role of $L_0 + \bL_0$ is replaced by dilatation,
while $L_0 - \bL_0$ is replaced by the ${d(d-1)\over 2}$ angular momentum
generators.

\subsection{Rotation algebra $\rightarrow$ Heisenberg algebra}
 
The isometries of $S^3$ also reduce to $SO(2) \times H(2)$. The
isometries of $S^3$ are $SU(2) \times SU(2)$ with generators $J_0,
J_\pm$ and $\bJ_0, \bJ_\pm$ respectively, with the usual algebra
\ben
[J_+,J_-]  =  2J_0~~~~~~~~[J_0, J_\pm] = \pm J_\pm 
\label{eq:csix}
\een
and similarly for the $\bJ$'s. Starting with a highest weight state 
$|j,\bj;0,0>$
\bea
J_+~|j,\bj;0,0>  & = & \bJ_+~|j,\bj;0,0>  = 0  \nn \\
J_0 ~|j,\bj;0,0> & = & j ~|j,\bj;0,0>~~~~\bJ_0 ~|j,\bj;0,0> = \bj ~|j,\bj;0,0>
\label{eq:cseven}
\eea
We create the other states in this representation as usual
\ben
|j,\bj,m,\bm> = (J_-)^m (\bJ_-)^\bm~|j,\bj;m,\bm>
\een
We are using an unconventional normalization for conveninence.
In this case the representation is finite dimensional, so that 
the maximum values of $m,\bm$ can be $2j,2\bj$ respectively.
Using the algebra we have well known relations
\bea
J_0 ~|j,\bj;m,\bm> & = & (j-m) ~|j,\bj,m,\bm>~~~~~\bJ_0 ~|j,\bj;m,\bm> = 
(\bj-\bm) ~|j,\bj,m,\bm> \nn \\
J_-~|j,\bj;m,\bm> & = & |j,\bj;m+1,\bm>~~~~~~~~~~\bJ_-~|j,\bj;m,\bm> 
= ~|j,\bj;m,\bm + 1> \nn \\
J_+ ~|j,\bj;m,\bm> & = & [2mj - m(m-1)]~|j,\bj;m-1,\bm> \nn \\
\bJ_+ ~|j,\bj;m,\bm>  & = &  [2\bm\bj - \bm(\bm-1)]~|j,\bj;m,\bm-1>
\label{eq:ceight}
\eea
When $j,\bj$ are large and $m << j, \bm << \bj$ 
we can ignore the fact that the
representations have finite dimensionalities $(2j+1, 2\bj+1)$. 
Furthermore, $J_-, \bJ_-$
and $J_+, \bJ_+$ become the creation and annihilation operators of two
commuting
Heisenberg algebras with planck's constants $2j$ and $2\bj$ respectively.
Taking sums and differences of $J_\pm$ and $\bJ_\pm$ we get a common
central charge for both the Heisenberg algebras which is $2(j+\bj)$,
while the difference $(j - \bj)$ is a $SO(2)$ angular momentum.

Once again the above considerations generalize to higher dimensional
rotation group $SO(d+2)$ leading to $SO(d) \times H(d)$.

\subsection{Combining the Conformal and Rotation algebras}

In the Penrose limit we are interested in states which have both large
conformal dimensions as well as large $SO(d+2)$ angular momenta. For
example in $d=2$ we are interested in states which are constructed 
out of highest weight states of the form
$|h,\bh;0,0> \otimes |j,\bj;0,0>$ with
\ben
h + \bh = j + \bj
\label{eq:wone}
\een
In a supersymmetric theory these states have a special significance 
since these are chiral primaries. The descendants of this state are 
\ben
(L_-)^n~(\bL_-)^\bn~(J_-)^m~(\bJ)^\bm ~|h,\bh;0,0> \otimes |j,\bj;0,0>
\een
Acting on the module, the product of conformal and rotation algebras
contract to $SO(2) \times SO(2) \times H(2) \times H(2)$ and the
central charges of the two $H(2)$'s are the same. To leading order
this common central charge is the value of the sum
$\half( L_0 + \bL_0 + J_0 + \bJ_0)$. The {\em difference}
$( L_0 + \bL_0) - (J_0 + \bJ_0)$ provides an outer automorphism of
the Heisenberg algebras. This is exactly what happens for the isometries
of the pp-wave.

\section{Heisenberg algebras in the CFT}

The discussion above has no reference to any specific way in which the
symmetries are realized. We have seen in the section 3 that
when these are realized as isometries of $AdS_3 \times S^3$ this
contraction corresponds to the Penrose limit which lead us to the
isometries of the pp-wave background.  We now want to see how this
mechanism happens in the CFT description.

In the CFT, the states form representations of the conformal and R
symmetry algebras. The pp-wave/CFT correspondence requires that a
subset of these states should survive in the large conformal weight
and large R charge limit, and should form representations of a product
of lower dimensional rotation and Heisenberg algebras. In this
section we will study this in a free field realization 
and show that this requirement implies that only a few of the original
angular momentum modes survive in this limit. We will perform the
analysis for $AdS_3 \times S^3$ and indicate how the result
generalizes to higher dimensions

The free field realization consists of four scalar fields living on
$S^1 \times {\tt time}$ which we will organize into two complex
scalars $U,V$ with their complex conjugates $\bU,\bV$. They together
form a vector representation of $SO(4)$. We will pick a particular
$SO(2)$ subgroup of $SO(4)$ such that $U$ has $R$ charge $+1$, $\bU$
has charge $-1$ and $V,\bV$ are neutral.

Each of the fields have a mode expansion
\bea
U & = & {i \over {\sqrt{4\pi}}}\sum_n {1\over n}[a_n e^{-inw} +
\ba_n e^{-in\bw}] \nn \\
\bU & = & {i \over {\sqrt{4\pi}}}\sum_n {1\over n}[\sa_n e^{-inw} +
\bsa_n e^{-in\bw}] \nn \\
V & = & {i \over {\sqrt{4\pi}}}\sum_n {1\over n}[b_n e^{-inw} +
\bb_n e^{-in\bw}] \nn \\
\bV & = & {i \over {\sqrt{4\pi}}}\sum_n {1\over n}[\sb_n e^{-inw} +
\bsb_n e^{-in\bw}]
\label{eq:rone}
\eea
where the nonvanishing commutators are
\ben
[ a_n, \sa_m ] = [ b_n , \sb_m ] =
[ \ba_n, \bsa_m ] = [ \bb_n , \bsb_m ] =
2n \delta_{m+n,0}
\label{eq:rtwo}
\een
In terms of these modes the generators $L_i, \bL_i$ of $SL(2,R)
\times SL(2,R)$ are given by
\ben
L_n = {1\over 2}\sum_{m=-\infty}^\infty (a_{n-m}\sa_m + b_{n-m}\sb_m)~~~~~~
(n = 0,\pm 1)
\label{eq:rthree}
\een
while the generators $J_i, \bJ_i$ of the $SU(2) \times SU(2) = SO(4)$
are given by
\bea
J_0 & = & {1\over 4} \int d\phi[ U{\delta \over \delta U}
- \bU{\delta \over \delta \bU} - V{\delta \over \delta V}
+ \bV{\delta \over \delta \bV}] \nn \\
J_+ & = & \half \int d\phi[ \bV{\delta \over \delta \bU}
-U{\delta \over \delta V}] \nn \\
J_- & = & \half \int d\phi[ \bU{\delta \over \delta \bV}
-V{\delta \over \delta U} ]\nn \\
\bJ_0 & = & {1\over 4} \int d\phi[ U{\delta \over \delta U}
- \bU{\delta \over \delta \bU} + V{\delta \over \delta V}
- \bV{\delta \over \delta \bV}] \nn \\
\bJ_+ & = & \half \int d\phi[ V{\delta \over \delta \bU}
-U{\delta \over \delta \bV}] \nn \\
\bJ_- & = & \half \int d\phi[ \bU{\delta \over \delta V}
-\bV{\delta \over \delta U}]
\label{eq:rfour}
\eea
The $SO(2)$ in question is generated by $J_0 + \bJ_0$. These may
be expressed in terms of modes using the mode expansions above.
All operators are assumed to be normal ordered.

The effect of the various $SL(2,R)$ generators on primary fields
is given by derivative operators as described in equation
(\ref{eq:thirteen}), while that the $SU(2)$ generators rotate the
fields among them.

To discuss states it is useful to euclideanize the time $\bt$ by
defining $\tau = - i\bt$ and perform a conformal transformation to
$R^2$ with complex coordinates $z,\bz$ 
\ben 
z = e^{\tau +
i\chi}~~~~~~~~\bz = e^{\tau - i\chi}
\label{eq;done}
\een 
On the plane the action of $L_i$ become
\bea
L_0 & =  -(z\partial_z + h) ~~~~~~& \bL_0  =  -(\bz\partial_\bz +\bh) \nn \\
L_- & =  -\partial_z ~~~~~~& \bL_- = -\partial_\bz     \nn \\
L_+ & =  -(z^2\partial_z + 2hz)~~~~~& L_+  =  -(\bz^2\partial_\bz + 2\bh\bz)
\label{eq:dtwo}
\eea
The $\bL$'s are obtained by replacing $z$ with $\bz$.

Consider the operator on the plane
\ben
\cO(z,\bz) = [\partial_z U (z) ]^j  [\partial_\bz U (\bz) ]^j 
\label{eq:rfive}
\een
At $z = \bz = 0$ this creates a highest weight state which we
will denote by 
\ben
|j;0,0;0,0> = \cO(0,0) |0>
\label{eq:rsix}
\een
The first slot simply labels the representation,
the next two zeros denote that they are at level $(0,0)$ of the
conformal algebra and at level $(0,0)$ of the rotation algebra.
(\ref{eq:rsix}) is clearly a highest weight state, killed by $L_+,\bL_+,
J_+, \bJ_+$. Furthermore
\ben
L_0 |j;0,0;0,0> = \bL_0 |j;0,0;0,0> = J_0 |j;0,0;0,0>
= \bJ_0 |j;0,0;0,0> = j~|j;0,0;0,0>
\een
In terms of the modes this state is
\ben
|j;0,0;0,0> = (a_{-1})^j (\ba_{-1})^j |0>
\label{eq:rseven}
\een
The descendants of this state are of the form
\ben
|j;n,\bn;m,\bm> = (L_-)^n (\bL_-)^\bn (J_-)^m (\bJ_-)^\bm
|j;0,0;0,0>
\label{eq:reight}
\een
In terms of the operator creating the state, action of $L_-$ is
simply a derivative $\partial_z$ while $J_-$ replaces one of the
fields $U$ by some other field.

From the discussion of the abstract algebras we know that 
\bea
L_+ |j;n,\bn;m,\bm> & = & [2jn + n(n-1)]~j;n-1,\bn;m,\bm> \nn \\
J_+ |j;n,\bn;m,\bm> & = & [2jm - m(m-1)]~|j;n,\bn;m,\bm>
\label{eq:rnine}
\eea
and similarly for the action of $\bL_+, \bJ_+$.
In the large $j$ and $n,\bn,m,\bm << j$ limit these relations
characterise Heisenberg algebras.
We want to see what is involved in this contraction.

\subsection{The conformal part}

Let us first discuss the conformal descendants. It is sufficient to
deal with one of the $SL(2,R)$ factors. The simplest nontrivial
example is
\ben
|j;2,0;0,0> = [j(j-1) (a_{-2})^2 (a_{-1})^{j-2}
+ 2j (a_{-3}) (a_{-1})^{j-1}]~(\ba_{-1})^j |j;0,0>
\label{eq:ddtwo}
\een
The operator which creates this state is
\ben
\partial_z^2 \cO (z) = [j(j-1) (\partial_z^2 U)^2 (\partial_z U
(z))^{j-2} + 2j (\partial_z^3 U) (\partial_z U (z))^{j-1}](\partial_\bz U)^j
\label{eq:ddone}
\een
the first term in (\ref{eq:ddtwo}) comes from the first term in
(\ref{eq:ddone}) and similarly for the second term. Even at this stage,
it is clear that the terms which involve $\partial^3$ is subdominant
in the large $j$ limit. Let us now check the action of $L_+$ on this state.
This may be easily calculated to yield

\ben
L_+ ~|j;2,0;0,0> = [4(j-1) + 6]~|j;1,0;0,0>
\label{eq:ddthree}
\een
which is consistent with the general result of (\ref{eq:cthree}).

The first term of the right hand side of (\ref{eq:ddthree}) comes from
the action of $L_+$ on the first term of (\ref{eq:ddtwo}), viz.  the
state $[j(j-1) (\alpha_{-2})^2
(\alpha_{-1})^{j-2}~|0>$. In the large $j$ small $n$ limit
only the first term in (\ref{eq:ddthree}) contributes and one has
\ben
L_+ ~|j;2,0;0,0> \simeq 4j ~|j;1,0;0,0>
\een
which, as argued above, is characteristic of Heisenberg algebra.

It is easy to generalize this result for arbitrary $n$.
$\partial_z^n$ distributes among the product of 
$(\partial_z U)$'s. Ignoring the factors of $\partial_\bz U$ which 
are always present,
\ben
\partial_z^n \cO (z) = [j(j-1)\cdots(j-n+1)](\partial
(\partial_z U))^n (\partial_z U)^{j-n} + \cdots
\label{eq:dnine}
\een

The elipses denote terms of the form $(\partial (\partial_z U))^{n-2}
(\partial^2 (\partial_z U)) (\partial_z U)^{j-n+1}$ and those with
more derivatives on a single $(\partial_z U)$.  Note that the first
term has $n$ powers of $j$ while the others have lower powers of
$j$. It may be easily seen that the state $|j:n,0;0,0>$ is of the form
\ben
|j;n,0;0,0> =
[j(j-1)\cdots(j-n+1)](a_{-2})^n~(a_{-1})^{j-n}
~|0> + O(j^{n-1})
\label{eq:dten}
\een
Once again the leading term in $L_+ ~|j;n,0;0,0>$ comes from this first term
in (\ref{eq:dten}) and leads to
\ben
L_+ ~|j;n,0;0,0> \simeq 2nj ~|j;n-1,0;0,0>
\label{eq:deleven}
\een
with subleading contributions which do not involve $j$.

It is now clear what is happening in the contraction of the conformal
algebra to the Heisenberg algebra. The raising operator $L_-$ acts as
a derivative on the complex plane. The only term which is relevant in
the action of $L_-^n$ on $(\partial_z U)^j$ is the term where we have
$n$ products of $(\partial_z^2 U)$ and none involving higher
derivatives on $U$. In terms of modes, this means that
the operators involve only $a_{-1}$ and $a_{-2}$. In the
theory on $S^1 \times {\tt time}$ these are the modes
with angular momentum $l=1$ and $l=2$ on the $S^1$.  Operators
constructed in this fashion automatically furnish a representation of
the Heisenberg algebra with a ``Planck's constant'' equal to $2j$.

It should be be possible to extend the above argument easily to
higher dimensions. For a CFT living in euclidean $R^d$ 
have $d$ raising operators in the conformal algebra. We start with
some elementary field $U$ with dimension $1$ and construct a
highest weight state by action of operators like $(U)^j$.
Descendants are obtained by action of these raising operators which
act like derivatives. In the large $j$ limit the only terms
which survive in these descendants are the ones which contain
single derivatives. There is one crucial difference from the two
dimensional case discussed above. Now the vacuum is not annihilated
by the zero mode of $(U)$. Thus the highest weight state has zero
angular momentum but nonzero energy (equal to $j$) and is created
by products of the zero mode of $(U)$. The descendants have higher
angular momenta. However the above discussion shows that in the large
$j$ limit only the $d$ states of lowest angular momentum, $l=1$
participate. This is precisely the proposal of BMN and plays an
important role in the dynamics at large R-charge \cite{larger}. We have
obtained this from the contraction of the conformal algebra to the
Heisenberg algebra.

\subsection{The R symmetry part}

The action of $J_-$ replaces one of the $U$'s in the highest weight
state by a $V$
\ben
[J_-, \cO] \sim j~[(\partial_z V)(\partial_z U)^{j-1}
(\partial_\bz U)^j + (\partial_\bz V)(\partial_z U)^{j}
(\partial_\bz U)^{j-1}]
\label{eq:rten}
\een 
The first level descendant of the R symmetry algebra is
\ben
|j;0,0;1,0> = J_- |j;0,0;0,0> = 
-{j \over 2}~[b_{-1} (a_{-1})^{j-1} (\ba_{-1})^j
+ \bb_{-1} (a_{-1})^{j} (\ba_{-1})^{j-1}]~|0>
\label{eq':releven}
\een
More powers of $J_-$ are going to bring in more of $V$'s
and $\bV$'s, but will not bring in any $\bU$'s. 

\noindent The lowest nontrivial state which contains a $\bU$ is
\bea
|j;0,0;1,1>  =   {j \over 2} &[&
(j-1) \sb_{-1}b_{-1} (a_{-1})^{j-2} (\ba_{-1})^j
+ j \sb_{-1}\bb_{-1}(a_{-1})^{j-1} (\ba_{-1})^{j -1} \nn \\
&-& \sa_{-1}(a_{-1})^{j-1} (\ba_{-1})^{j} \nn \\
&+& (j-1) \bsb_{-1}\bb_{-1} (\ba_{-1})^{j-2} (a_{-1})^j
+ j \bsb_{-1}b_{-1}(\ba_{-1})^{j-1} (a_{-1})^{j -1} \nn \\
&-& \bsa_{-1}(\ba_{-1})^{j-1} (a_{-1})^{j}]~|0>
\label{eq:rthirteen}
\eea

\noindent The operator structure for these terms are
\bea
|j;0,0;1,1>  =   {j \over 2} &[&
(j-1) (\pz \bV) (\pz V) (\pz U)^{j-2} (\pbz U)^j
+ j (\pz \bV) (\pbz V) (\pz U)^{j-1} (\pbz U)^{j-1} \nn \\
&-& (\pz \bU)(\pz U)^{j-1} (\pbz U)^{j} \nn \\
&+& (j-1) (\pbz \bV) (\pbz V) (\pbz U)^{j-2} (\pz U)^j
+ j (\pbz \bV) (\pz V) (\pbz U)^{j-1} (\pz U)^{j-1} \nn \\
&-&  (\pbz \bU)(\pbz U)^{j-1} (\pz U)^{j}] 
\label{eq:rfourteen}
\eea
The term which involves $\bU$ contains
lower powers of $j$ and would be subdominant. This becomes
clear when we consider the action of $J_+$ on this state
and see how it acts as an annihilation operator of a
Heisenberg algebra in the large $j$ limit. One gets
\bea
J_+ |j;0,0;1,1>  =  - {j \over 2} &[&
(j-1) b_{-1} (a_{-1})^{j-1} (\ba_{-1})^j + 
j \bb_{-1} (a_{-1})^{j} (\ba_{-1})^{j-1} \nn \\
&+& b_{-1} (a_{-1})^{j-1} (\ba_{-1})^j \nn \\
&+& (j-1)\bb_{-1} (a_{-1})^{j} (\ba_{-1})^{j-1}
+ j  b_{-1} (a_{-1})^{j-1} (\ba_{-1})^j \nn \\
&+& b_{-1} (a_{-1})^{j} (\ba_{-1})^{j-1}]~|0>
\label{eq:rfifteen}
\eea
Each term in (\ref{eq:rfifteen}) is the action of $J_+$ on
the corresponding term in (\ref{eq:rfourteen}). The final
result is as expected
\ben
J_+ |j;0,0;1,1> = 2j |j;0,0;0,1>
\een
At this level, the action of $J_+$ 
is like that of a Heisenberg annihilation operator
even for finite $j$.  However the significant point
is that in the large $j$ limit the terms which contain $\sa_{-1}$ in
(\ref{eq:rfourteen}) do not contribute. The situation becomes clear at
the next level, viz. the action of $J_+$ on $|j;0,0;2,1>$. Following
similar manipulations we find that the corresponding operator which
contains $\bU$ does not contribute in the large $j$ limit.

The basic reason behind this may be found by examining the charges,
equation (\ref{eq:rfour}). It is clear from these expressions that
$\bU$ can be introduced only if there is a $V$ or a $\bV$ in the
operator in an earlier stage, which may be replaced by $\bU$.  However
each such term is accompanied by a term which replaces one of the
$U$'s by a $V$ or a $\bV$ - and there are lot more of these terms
since the number of $V,\bV$ are always much smaller than the number of
$U$'s.

We therefore conclude that the contraction of the rotation algebra to
a Heisenberg algebra shows that the states which survive in this limit
do not contain modes of the field $\bU$.

This argument may be generalized to higher dimensions trivially.  This
is because we are working with an internal symmetry. Once again the
major difference is that the fields $U,V$ themselves are conformal
fields and therefore the highest weight states are created by $U^j$
itself. Consequently all the modes in this subsection would be modes
with zero angular momentum rather than with angular momenta $\pm
1$. Action of R symmetry generators of course do not change the
angular momentum.

\subsection{Combining the conformal and R symmetry parts}

Let us now consider mixed descendants. The lowest such state is
\bea
|j;1,0;1,0> = -j &[& (j-1) a_{-2} b_{-1} (a_{-1})^{j-2}
(\ba_{-1})^j + j a_{-2} \bb_{-1} (a_{-1})^{j-1}
(\ba_{-1})^{j-1}  \nn \\
&+& b_{-2} b_{-1}(a_{-1})^{j-1}
(\ba_{-1})^{j}]~|0>
\label{eq:sone}
\eea
the structure of the operator which creates this state is
\bea
-j &[& (j-1) (\pz^2 U)(\pz V)(\pz U)^{j-2}(\pbz U)^j
+ j  (\pz^2 U)(\pbz V)(\pz U)^{j-1}(\pbz U)^{j-1} \nn \\
&+& (\pz^2 V)(\pz V) (\pz U)^{j-1}(\pbz U)^{j}]
\label{eq:stwo}
\eea
This is the first nontrivial operator which contains $\pz^2 V$.
However it is clear that the term which involves $\pz V$ is
subdominant in the large $j$ limit. One can now go ahead
and examine the action of $J_+$ or $L_+$ on this state and
verify that the expected answers follow. In all these
steps the term which involves $\pz^2 V$ does not contribute
in the large $j$ limit. A similar result follows for 
$\partial_z^2 \bV$ when we consider the state $|j;1,0;0,1>$.
The situation becomes clearer at 
the next level, which we have checked, but will not present
here.

Once again the reason behind this is clear from the generators.
Higher angular momentum modes of $V,\bV$ can be obtained only
when $L_-$ or $\bL_-$ act. These are derivatives which get
distributed over all the terms in the product, and since there
are always lot more $U$'s compared to $V$ or $\bV$'s, the
dominant terms are those in which the derivatives act on $U$'s.

We therefore conclude that in this limit the fields $V$ must
be restricted to its lowest nontrivial angular momentum mode.
Once again the argument generalizes to higher dimensions
with the obvious differences noted above.

Let us therefore list the various facts which result from
examining the contraction of the product of conformal and
R symmetry groups to lower dimensional rotation and R symmetry
groups and Heisenberg groups. 
Starting with a state constructed
from products of an elementary field $U$ with $SO(2)$ R charge $+1$

\begin{enumerate}

\item The field $U$ is restricted to its two lowest nontrivial angular
momentum states. (For $d > 2$ these are the zero mode and
the $d$ $l=1$ modes.)

\item The field $\bU$ does not appear in the operators which
create the states

\item The fields which are neutral under the $SO(2)$ are
restricted to their lowest nontrivial angular momentum state.

\end{enumerate}

These are crucial ingredients in the work of BMN \cite{bmn}. We
have obtained them from symmetry considerations.

\subsection{States in the dual gauge theory}

For the case of most interest, $AdS_5 \times S^5$ the dual CFT is a
${\cal N}=4, ~SU(N)$ gauge theory living on $S^3 \times {\tt time}$.
The conformal group is $SO(4,2)$ and the R-symmetry is $SO(6)$. In the
pp-wave description, the Penrose limit leads to $SO(4) \times SO(4)
\times H(4) \times H(4)$ and according to the proposal of BMN is the
states of string theory in the pp-wave background are a subset of
states in the gauge theory with large R-charge $J$ and large dimension
$\Delta$ and small $\Delta - J$.

The considerations of the above subsections may be viewed as a
carricature of this. We have not considered the effects of the gauge
group and our considerations were restricted to free field theory.

The contraction of the conformal and the R-symmetry algebras does not
depend on the fact that these symmetries act on gauge fields and
adjoint scalars. The specific way these are represented, however,
differ from the
considerations of this section - the operators in question involve
traces over the gauge group. Nevertheless we expect that the
conclusions remain at the free field level.  The essential point is
that the operators which dominate in the large $j$ limit are those
which are obtained when the lowering operators act on the string of
$U$'s rather than other fields which are brought down at earlier
levels. This should lead to a restricted Hilbert space ${\cal H}_h
\subset {\cal H}$
of states created by operators which do not contain higher angular
modes of the elementary fields.

When we turn on interactions, one has to ask about transition
amplitudes between states in the restricted hilbert state ${\cal H}_h$
and the states in ${\cal H} - {\cal H}_h$.  In \cite{bmn} it was
argued that at large $N$ operators containing higher angular momentum
modes acquire a large anomalous dimension and therefore decoouple from
the theory, so that these transition amplitudes vanish. This is
where large N and large 't Hooft coupling enters into the discussion
and is the reason why our conclusions regarding a contraction of the
Hilbert space would be valid in the full theory.

Usually the large N limit reduces the allowed interactions to planar
diagrams but do not reduce the free Hilbert space. On the other hand,
as we have seen above, large conformal weight and large R-charge
reduce the free Hilbert space without affecting the intensity of
interaction. When we combine the two together we get a consistent
truncation of the Hilbert space. Supersymmetry is a crucial ingredient
in this since this ensures that amplitudes for transitions between
states in the reduced space remain finite while those which take us
out of this reduced space are suppressed. For nonsupersymmetric
theories like QCD, very likely the analog of the large $J$ limit
contraction should be defined as an infinite momentum frame limit
similar to the original definition of Matrix theory as the infinite
momentum frame limit of 11 dimensional supergravity.

\subsection{A connection to the bulk}

The bulk generators of the Heisenberg algebra, equation
(\ref{eq:bthree}) appear to be quite different from the representation
in conformal field theory discussed in the previous section. Unlike
$AdS \times S$ spacetimes the bulk generators do not reduce to the CFT
generators when restricted to some region of the spacetime (e.g. the
boundary). Is there any sense in which the CFT continues to provide
a ``holographic'' description of the bulk ?

In $AdS \times S$, the holographic representation of the $AdS$
isometries is in terms of conformal symmetries of a CFT while the $S$
isometries are represented as internal symmetries. When we take the
Penrose limit both of these isometries becomes products of rotation
group and Heisenberg group and there is a $Z_2$ symmetry between
these. One would expect that the same phenomenon appears in a
holographic description.

The above discussion shows that the CFT essentially reduces to a
finite number of quantum mechanical degrees of freedom. For $d > 2$
these include the $d+1$ zero modes of the scalars and $d$ lowest
angular momentum modes of the scalar $U$.  The symmetries may the then
written in terms of these modes and derivatives with respect to them.
The generators of the Heisenberg algebra in the bulk involve
derivatives with respect to the transverse coordinates to the
pp-wave. In \cite{dgr} it was proposed that this implies that there
should a holographic representation in $d$ euclidean dimensions. In
view of the above results it is tempting to attempt a different
interpretation, viz. the transverse coordinates in the bulk should be
thought of as these $2d$ quantum mechanical degrees of freedom. It
appears, however that the CFT furnishes a Bargmann-Fock representation
of the harmonic oscillator algebra while the bulk furnishes a usual
coordinate representation \footnote{ A related point has been
made in Arutyunov and Sokatchev, \cite{symmetries}}.  
This would make the $Z_2$ symmetry more
manifest.  The details of this correspondence, especially the role of
the gauge field remain to be understood.

\section{Correlators in the large R charge limit}

In the standard $AdS/CFT$ correspondence the n-point correlators 
of conformal fields are related to bulk correlators via boundary-to-bulk 
Green's functions \cite{adscft}. Consider for example the two point
function of an operator $\cO_{h,n_i}$ which has vanishing anomalous
dimension. The lowest energy state in the CFT
defined on $S^{d-1}$ created by this operator are the ones which are
obtained from a primary state by the action of $d$ raising operators
$n_i$ times. The energy
of this state is $h + \sum_i n_i$. This is also the conformal dimension
of the operator, so that on $R^d$ the two point function is given by
\ben
< \cO_{h,n_i} (x) \cO_{h,n_i} (0) > = {1\over |x|^{2h + 2\cn}}
\label{eq:qone}
\een
where $\cn = \sum_i n_i$. Here $|x|$ denotes a radial coordinate in the
$R^d$. This is related to the euclidean time coordinate $\tau$ of the CFT
defined on a $S^{d-1}$ by $|x| = e^{\tau}$. Since this is the global 
time in the $AdS$ we get the relationship
\ben
<h,n_i | e^{-\tau H} |h, n_i> = {1\over |x|^{2h + 2\cn}}
\label{eq:qtwo}
\een
where $H$ is the hamiltonian of the bulk theory.

In the strict $h = \infty $ limit the above correlator is zero for 
$|x| > 1 $  and infinite for $|x| < 1$. To make concrete connections
to bulk quantities it is first necessary to define finite quantities
starting from these correlation functions. 

We have seen that in the Penrose limit the conformal and R symmetry algebras 
reduce
to Heisenberg algebras and we can write
\ben
H = C + P
\een
where $C$ denotes the common central extension and $P$ denotes the
outer automorphism. In the bulk we have $C \sim p_- = p^+$ and
$P = p_+ = p^-$. This suggests that we {\em define} normalized 
two point functions 
\ben
G(|x|)  = 
{<h,n_i | e^{-\tau H} |h, n_i> \over <h,0 | e^{-\tau H} |h, 0>}
\een
Note that in the bulk 
the state $|h,0>$ is the light cone vacuum with $p_- = h$
Then the above correspondence implies
\ben\label{pro}
G(|x|) = <h,n_i|e^{-\tau P^-}|h,n_i>
\een

We will now consider the large $J$ limit 
of CFT three point functions. As is well known, in CFT the three point
function is completely determined by the OPE
\ben
O_{i}(x)O_{j}(0) = C_{ijk} \frac{1}{|x|^{(h_{k}-h_{i}-h_{j})}} O_{k}(x) 
\een
Let us now assume that the conformal weights are
such that $h_{i}=J_{i} + p_{i}$ for $p_{i}$ small but finite numbers
and where $J_{i}$ represent $U(1)$ charges. We
are interested in the limit of the OPE for large $J_{i}$. Conservation 
of $U(1)$ charge $J_{k} = J_{i} + J_{j}$ implies that the space
time dependence of the OPE is simply $\frac{1}{|x|^{(p_{k}-p_{i}-p_{j})}}$ i.e
it is independent of $J$. However the structure constants $C_{ijk}$ will
generically depend on the values of the $U(1)$ charges $J_{i}$. For instance
in the case of BMN operators we have 
$C_{ijk} = \frac{\sqrt{J_{i}J_{j}J_{k}}}{N}$. In the BMN limit where
$J \sim \sqrt{N}$, the structure constant scale like 
$1/{\sqrt{J}}$ and therefore goes to zero in the limit $J = \infty$. 
This motivates a redefinition of the structure constants to have
a finite OPE in the 
large $J$ limit.
In what follows we suggest a way to do it. Let
us first write the OPE as
\ben
C_{ijk}~ [1 - (p_{k}-p_{i}-p_{j})~\log|x|]~O_{k}(0)
\een
and let us define a finite double limit, $J \rightarrow \infty$
and $|x| \rightarrow 0$ with $y = \log |x| / {\sqrt{J}}$ held finite.
In other words as we increase $J$ we focus on smaller values
of $|x|$. Now we have
\ben
C_{ijk}~\log|x|= \frac{a^{2}}{\sqrt{J}} \log(|x|)
\een
where $a^2 = J^2/N$.
The OPE can be now represented
as a vertex operator, namely
\ben\label{vertex}
<i,j,k|V> = y.a^{2}.(p_{k}-p_{i}-p_{j})
\een
with the new ``blow up'' variable $y$ representing
now the interaction amplitude. This structure of the vertex operator
was already suggested in ref \cite{constable}.
In summary, we simply observe that in the large $J$ limit the main
contribution to the OPE comes from the contact terms appearing in the
limit $|x-y| = 0$ with the corresponding pole being compensated by the
zero of the structure constant in the large $J$ limit.  It is natural
to conjecture that in the large $J$ limit generic correlators could be
mapped into string like diagrams defined in terms of the vertex
opertor (\ref{vertex}) and the free propagators (\ref{pro}). For
recent discussions on the three point function see references
\cite{bnas,threepoint}.

\section{Conclusions}

We have studied how the conformal and R-symmetry algebras in a CFT
contract to products of lower dimensional rotation and R symmetry
algebras and Heisenberg algebras. This parallels the similar
contraction in the Penrose limit of $AdS \times S$ spacetimes. We have
argued that in this contraction process, higher angular momentum modes
of fields in the CFT decouple and either the field with $SO(2)$
R-charge $+1$ or with $SO(2)$ R-charge $-1$ remain (but not both).
We have not considered the supersymmetries; we expect that analogous
considerations for the superconformal algebras will provide more
insight.

The fact that at large weight and large R charge one gets Heisenberg
algebras is quite general. We have explicitly worked out how this
happens for the $d=2$ case, but it is clear that the result
generalizes to any number of dimensions. The reduction of the Hilbert
space has been, however, demonstrated in a toy model of $1+1$
dimensional CFT which is a recognizable carricature of higher
dimensional models. This makes it quite plausible that one could
generalize the considerations to higher dimensions. However several
crucial ingredients are absent in our toy model - these are related to
the fact that in higher dimensions the CFT is a gauge theory and one
has to perform a large N limit of this gauge theory. It is important
to investigate whether such symmetry considerations may be used to
understand the phenomenon in gauge theory - this would also elucidate
the role of large $N$ limit. The intricacies of operator structure
and operator mixing \cite{operators} would be relevant to this.

In this paper we have dealt with supergravity modes in the bulk and
their CFT descriptions. These are the states which are created using
the isometries of the geometry. The most interesting feature of the
pp wave - CFT correspondence is, however the fact that a string theory
is tractable in this geometry and \cite{bmn} have found how to describe
the higher stringy modes in the gauge theory. We have not dealt with 
such stringy modes in this paper. However it is reasonable to believe
that one can understand the spectroscopy of stringy modes in terms of
worldsheet current algebras and one has to understand
how this is realized in the CFT. A requirement that the CFT encodes
this current algebra correctly should throw light on the proposal
of \cite{bmn} for CFT description of stringy states.

Finally, the meaning of holography in this correspondence remains
unclear. The present paper reinforces the claim that a large R-charge
limit of the original CFT is dual to the bulk theory. But in what
sense is this a holographic description ? The dual CFT lives on a
$S^{d-1} \times {\tt time}$ which is the boundary of the $AdS_{d+1}$
spacetime {\em before} any Penrose limit. The Penrose limit, on the
contrary, focusses on the deep interior of the $AdS$ which is not a
part of the pp-wave geometry. The fact that the CFT essentially
becomes a quantum mechanical system might suggest, however, that it is
natural to place this on the one dimensional boundary of pp-wave - as
suggested in \cite{bnas}. However, we do not have any concrete check
of this idea yet. What does appear to be true is that all the
symmetries of the bulk are realized as internal symmetries of this
effective quantum mechanics. It appears likely that for
the 11 dimensional pp-wave the theory on the boundary is the 
Matrix theory in this background written down in \cite{bmn} and
further studied in \cite{matrix}. It would be interesting to see
if this can be made concrete.

The key point to understand is whether
the scale of the CFT appears as some coordinate in the pp-wave in a
way similar to what happened in the AdS/CFT correspondence. One
possible scenario is that {\em worldsheet dilatations} become a
holographic coordinate, in a way similar to noncritical string theory
\cite{dnw,pola}. The CFT would be then similar to
the $c=1$ matrix model \cite{dj}.  If this is true, one should be able
to recast the contracted form of the beta function equations of the
CFT as worldsheet beta functions which would be related to bulk
equations of motion in the usual way. For a recent discussion of
RG flows in this context see \cite{rg}.

\section{Acknowledgements}

S.R.D. would like to thank S. Mathur, S. Minwalla, Y. Nambu and
especially A. Jevicki for discussions. 
C.G. would like to thank A. Uranga for discussions.
The work of S.R.D. is partially
supported by U.S. DOE contract DE-FG01-00ER45832. The work of C.G. is
partially supported by grant AEN2000-1584.


\begin{thebibliography}{99}

\bibitem{metsaev} R. Metsaev, {\em Nucl. Phys.} {\bf B 625} (2002) 70,
{\tt hep-th/0112044}; R. Metsaev and A. Tseytlin, {\tt hep-th/0202109};
J. Russo and A. Tseytlin, {\it JHEP} {\bf 0204} (2002) 021, 
{\tt  hep-th/0202179}. 

\bibitem{penrose} R. Penrose, in {\em Differential geometry and Relativity},
Riedel, Dordrecht, 1976;
R. G\"uven, {\it Phys. Lett.} {\bf B482} (2000) 255,
{\tt hep-th/0005061};
J. Kowalski-Glikman, {\it Phys. Lett.} {\bf 134B}
(1984) 194.

\bibitem{blau} M. Blau, J. Figueroa-O'Farrill, C. Hull and
G. Papadopoulos, {\em JHEP} {\bf 0201} (2002) 047, {\tt
hep-th/0201081} and {\tt hep-th/0110242}; M. Blau,
J. Figueroa-O'Farrill and G. Papadopoulos, {\tt
hep-th/0202111};
J. Kowalski-Glikman, {\it Phys. Lett.} {\bf 150B}
(1985) 125; P. Meesen, {\tt hep-th/0111031}.

\bibitem{bmn} D. Berenstein, J. Maldacena and H. Nastase, {\tt
hep-th/0202021}

\bibitem{dgr} S.R. Das, C. Gomez and S.J. Rey, {\tt hep-th/0203164}.

\bibitem{kp} E. Kiritsis and B. Pioline, {\tt hep-th/0204004}.

\bibitem{lor} R. Leigh, K. Okuyama and M. Rozali, {\tt hep-th/0204026}.

\bibitem{bnas} D. Berenstein and H. Nastase, {\tt hep-th/0205048}.

\bibitem{siopsis} G. Siopsis, {\tt hep-th/0205302}.

\bibitem{other}N. Itzhaki, I. Klebanov and S. Mukhi {\it JHEP}
0203:048,2002, {\tt hep-th/0202153}; J. Gomis and H. Ooguri,
{\tt hep-th/0202157}; L.P. Zayas and J. Sonnenschein, {\it JHEP 0205}
(2002) 010, {\tt hep-th/0202186};M. Alishahiha and
M.M. Sheikh-Jabbari, {\it Phys. Lett.}  {\bf B535} (2002) 328, {\tt
hep-th/0203018}; N. Kim, A. Pankiewicz, S.J. Rey, and S. Theisen, {\tt
hep-th/0203080}; M. Cvetic, H. Lu and C. Pope, {\tt hep-th/0203082};
T. Takayanagi and S. Terashima, {\tt hep-th/0203093};
U. Gursoy, C. Nunez and M. Schvellinger, {\tt hep-th/0203124};
E. Floratos and A. Kehagias, {\tt hep-th/0203134};
J. Michelson, {\tt hep-th/0203140}; 
M. Cvetic, H.Lu and C. Pope, {\tt hep-th/0203229};
J. Gauntlett and C. Hull , {\tt hep-th/0203255};
H. Lu and J. Vazquez-Portiz, {\tt hep-th/0204001};
S. Mukhi, M. Rangamani and E. Verlinde, {\tt hep-th/0204147};
I. Bakas and K. Sfetsos, {\tt hep-th/0205006};
K. Oh and R. Tatar, {\tt hep-th/0205067};
C. Ahn, {\tt hep-th/0205008}, {\tt hep-th/0205109} and {\tt hep-th/0206029};
Y. Hikida and Y. Sugawara, {\tt hep-th/0205200};
V. Hubeny, M. Rangamani and E. Verlinde, {\tt hep-th/0205258};
S. Seki, {\tt hep-th/0205266};
D. Mateos and S. Ng, {\tt hep-th/0205291};
F. Bigazzi, A. Cotrone, L. Girardello and A. Zaffaroni {\tt hep-th/0205296}.

\bibitem{dbranes} M. Billo and I. Pesando, {\tt hep-th/0203028};
C. Chu and P. Ho, {\tt hep-th/0203186};
A. Dabholkar and S. Parvizi, {\tt hep-th/0203231};
D. Berenstein, E. gava, J. Maldacena, K. Narain and H. Nastase,
{\tt hep-th/0203249};
P. Lee and J. Park, {\tt hep-th/0203257};
A. Kumar, R. Nayak and Sanjay, {\tt hep-th/0204025};
D. Bak, {\tt hep-th/0204033};
K. Skenderis and M. Taylor, {\tt hep-th/0204054};
V. Balasubramanian, M. Huang, T. levi and A. Naqvi, {\tt hep-th/0204196};
H. Takanayagi and T. Takanayagi, {\it JHEP} {\bf 0205} (2002) 012,
{\tt hep-th/0204234};
H. Singh, {\tt hep-th/0205020};
P. Bain, P. Meesen and M. Zamaklar, {\tt hep-th/0205106};
M. Alishalia and A. Kumar, {\tt hep-th/0205134};
O. Bergman, M. Gaberdiel and M. Green, {\tt hep-th/0205183};
B. janssen and Y. Lozano, {\tt hep-th/0205254};
S. Pal, {\tt hep-th/0205303}

\bibitem{semiclassical} S. Gubser, I. Klebanov and A. Polyakov,
{\tt hep-th/0204051};
S. Frolov and A. Tseytlin, {\tt hep-th/0204226};
J. Russo, {\tt hep-th/0205244};
A. Armoni, J. Barbon and A. Petkou, {\tt hep-th/0205280};

\bibitem{strings}M. Spradlin and A. Volovich, {\tt hep-th/0204146};
M. Alishalia and M. Sheikh-Jabbari, {\tt hep-th/0204174};
A. Parnachev and D. Sahakyan, {\tt hep-th/0205015};
Y. Imamura, {\tt hep-th/0204200};
C. Chu, P. Ho and F.L. Lin, {\tt hep-th/0205218};
T. Takanayagi, {\tt hep-th/0206010}

\bibitem{larger} C. Kristjansen, J. Plefka, G. Semenoff and M. Staudacher,
{\tt hep-th/0205033}; D. Gross, A. Mikhailov and R. Roiban,
{\tt hep-th/0205066}; N.R. Constable, D.Z. Freedman, M. Headrick,
S. Minwalla, L. Motl, A. Postnikov and W. Skiba, {\tt hep-th/0205089}.

\bibitem{bh} M. Li, {\tt hep-th/0205043};
S. Mathur, A. Saxena and Y. Srivastava, {\tt hep-th/0205136}

\bibitem{symmetries} R. Metsaev and A. Tseytlin in
\cite{metsaev}; M. Hatsuda, K. Kamimura and M. Sakaguchi,
{\tt hep-th/0202190}; G. Arutyunov and E. Sokatchev,
{\tt hep-th/0205270}.

\bibitem{adscft} See O. Aharony, S. Gubser, J. Maldacena, H. Ooguri and
Y. Oz, {\it Phys.Rept.} {\bf 323} (2000) 183, {\tt hep-th/9905111.}
and references to original literature.

\bibitem{bala} V. Balasubramanian, P. Kraus and A. Lawrence, {\em Phys. Rev.}
{\bf D 59} (1999) 046003.

\bibitem{constable} N.R. Constable et.al. in \cite{larger}

\bibitem{threepoint} D. Berenstein and H. Nastase
in \cite{bnas}; Y. Kiem, Y. Kim, S. Lee and J. Park, {\tt
hep-th/0205279.}; M. Huang, {\tt hep-th/0205311.}; C.S. Chu,
V. V. Khoze and G. Travaglini, {\tt hep-th/0206005.}

\bibitem{operators} S. Corley, A. Jevicki and S. Ramgoolam,
{\tt hep-th/0201222};
S. Corley and S. Ramgoolam, {\tt hep-th/0205221};
M. Bianchi, B. Eden, G. Rossi and Y. Stanev, {\tt hep-th/0205321};
G. Arutynov, S. Penati, A. Petkou, A. Santabrogio and E. Sokatchev,
{\tt hep-th/0206020}.

\bibitem{matrix} R. Gopakumar, {\tt hep-th/0205174};
K. Dasgupta, M. Sheikh-Jabbari and M. van Raamsdonk, {\tt hep-th/0205185};
G. Bonelli, {\tt hep-th/0205213}


\bibitem{dnw} S.R. Das, S. Naik and S.R. Wadia,
{\em Mod. Phys. Lett.} {\bf A4} (1989) 1033; S.R. Das, A. Dhar
and S.R. Wadia, {\em Mod. Phys. Lett.} {\bf A5} (1990) 799;
A. Dhar and S.R. Wadia, {\em Nucl.Phys.} {\bf B590} (2000) 261, 
{\tt hep-th/0006043}. 

\bibitem{pola} A.M. Polyakov, {\em Proceedings of Les Houches School,
1993}; A.M. Polyakov, {\em Nucl.Phys.Proc.Suppl.}  {\bf 68} (1998) 1,
{\tt hep-th/9711002.}; A.M. Polyakov, {\em Int.J.Mod.Phys.}  {\bf A14}
(1999) 645, {\tt hep-th/9809057}; S. Gubser, I. Klebanov and
A. Polyakov, {\em Phys.Lett.} {\bf B428} (1998) 105.

\bibitem{dj} S.R. Das and A. Jevicki, {\em Mod. Phys. Lett.}
{\bf A5} (1990) 1639.

\bibitem{rg} E. Gimon, L.A. Pando Zayas and J. Sonnenschein,
{\tt hep-th/0206033}


\end{thebibliography}
\end{document}